\documentclass[aps,pra,reprint]{revtex4-1}
\usepackage{graphicx}
\usepackage{times}
\usepackage{bm}
\usepackage{amsmath,amssymb}

\begin{document}
\title{Angular correlations in two-photon spectroscopy of hydrogen}

\author{A. Anikin}
\affiliation{ 
Department of Physics, St. Petersburg State University, Petrodvorets, Oulianovskaya 1, 198504, St. Petersburg, Russia
}

\author{T. Zalialiutdinov}
\affiliation{ 
Department of Physics, St. Petersburg State University, Petrodvorets, Oulianovskaya 1, 198504, St. Petersburg, Russia
}

\author{D. Solovyev}
\affiliation{ 
Department of Physics, St. Petersburg State University, Petrodvorets, Oulianovskaya 1, 198504, St. Petersburg, Russia
}

\email[E-mail:]{t.zalialiutdinov@spbu.ru}

\begin{abstract}
In the present paper, we consider nonresonant corrections to $ 2s-ns/nd $ transition frequencies in hydrogen for the experiments based on two-photon spectroscopy. A detailed study of angular correlations of quantum interference effects within the framework of rigorous quantum electrodynamics is given. Closed expressions for the resonant two-photon scattering cross sections on an atomic level with dependence on all atomic quantum numbers including fine and hyperfine structure are derived. These expressions are applied for the description of experiments based on two-photon spectroscopy with fixing of incident (outgoing) photon polarizations. We demonstrate that for certain experimental geometry nonresonant corrections could be significant for the determination of Rydberg constant and proton charge radius.
\end{abstract}

\maketitle

\section{Introduction}
Over the past two decades the absolute measurement of transition frequencies of hydrogen has been continuously improved with the aim of determining the Rydberg constant and testing quantum electrodynamics calculations. Recently in \cite{beyer} the accuracy of experiment on the determination of $ 2s-4p $ transition frequency achieved the level of few kHz where the analysis quantum interference effects (QIE) between states separated by the fine structure interval played a decisive role. However, in order to reliably determine the fundamental constants, it is also necessary to measure the frequencies of other transitions in hydrogen \cite{Karshenboim, LambShift-H, science2020, solovyev2018problem}. 

In the present paper we focus on the analysis of QIE in the measurements of $ 2s\rightarrow ns/nd $ ($ n=4,\;6,\;8,\;12 $ is the principal quantum number) transition frequencies. In these experiments hydrogen atoms in atomic beam are prepared in $ 2s_{1/2}^{F=1} $ state and excited to the $ ns_{1/2}^{F=1} $ or $ nd_{3/2}^{F=2} $ state by the absorption of two polarized laser photons propagating in opposite direction \cite{Schwob, deB-0,deB-1,deB-2}. Then the detection of the excited $ nd$ fraction of atoms could be observed via its fluorescence (i.e. decay to the $ 2p $ state) \cite{Weitz} or the decrease in metastable $ 2s $ signal \cite{Weitz-1,Nez}. In both cases the interference between different fine sublevels $ nd_{3/2}^{F=2} $ and $ nd_{5/2}^{F=2} $ occurs, which leads to the asymmetry of line profile. 

In \cite{Weitz} it was shown that the detection of the excited $ ns/nd$ fraction of atoms via its fluorescence has a considerably higher potential accuracy than the experiments monitoring the rate of metastable state quenching \cite{deB-1,deB-2}, which were limited by a large background of unexcited $ 2s $ atoms. Recently, in \cite{solovyev2020proton} it was shown that nonresonant (NR) correction to the $ 2s_{1/2}^{F=1}-nd^{F=2}_{3/2(5/2)} $ transition frequencies measured in the experiments of the type \cite{deB-1,deB-2} reaches the level of several kHz, making it significant in determining the proton charge radius $ r_{p} $ and Rydberg constant $ R_{\infty} $. In particular, taking into account for NR correction the averaged values of $ r_{p} $ were found as coinciding with the resuts reported in \cite{beyer}.

This paper is devoted to the theoretical description of experimental method based on the detection of fluorescence of outgoing photons. In the experiments with one-photon scattering, \cite{Jent-Mohr,S-2020-importance,Amaro-2015} it was shown that for the certain geometry the influence of NR effects can be significantly reduced. Extending approach developed in \cite{S-2020-importance} for the analysis of angular correlations of NR corrections we derive expressions for the cross section of resonant two-photon scattering on hydrogen atom levels taking into account the fine and hyperfine structure. These expressions are dependent on the directions and polarizations of the incident (absorbed) and outgoing (emitted) photons. They allow one to describe different correlations between directions and polarizations of three photons. The results of evaluation are then applied to the derivation of NR corrections to the two-photon scattering cross sections and transition frequencies.

The paper is organized as follows. In section \ref{section1} within the rigorous quantum electrodynamics (QED) and S-matrix formalism, we derive the cross section for resonant two-photon scattering on the atomic electron with account for the fine and hyperfine level structure. The application of derived expressions is considered on the example of $ 2s\rightarrow ns $ and $ 2s\rightarrow nd $  two-photon transitions in sections \ref{section2} and \ref{section3} respectively.  The analysis of angular correlations is given in section \ref{conclusion}. Details of mathematical derivations are set out in the Appendix A. The general expressions for NR corrections to transition frequencies are given in Appendix B. The relativistic units $ \hbar=c=1 $ are used throughout the paper. 

\section{QED theory of resonant two-photon scattering on atomic electron}
\label{section1}

For an accurate description of NR corrections to the atomic transition frequencies, it is natural to employ the QED theory of atomic processes developed in particular in \cite{Andr,ZSLP-report}. The resonant scattering of two photons corresponds to the case when the sum of frequencies of incident photons are equal to the difference of atomic level energies $ \omega = E_{n}-E_{i} $ for the particular $ n $ state. Denoting by $i,\,n,\,f$ the initial, intermediate, and final atomic
states, respectively, we assume the standard set of quantum numbers taking into account fine and hyperfine structure of levels: principal quantum number $n$, electron orbital angular momentum $l$, electron total electron momentum $j$, atomic angular momentum $F$, and its projection $M_{F}$.

According to the resonant approximation, \cite{Andr,ZSLP-report} the only one term for the chosen resonant $  n$ level can be retained in the sum over intermediate states in the scattering amplitude. Then arising divergent contribution should be regularized by an inclusion infinite set of Feynman graphs representing the self-energy of the bound electron. This leads to the arrival of the level width $ \Gamma_{n} $ in the energy denominator. In the lowest order, the regularized in this way resonant contribution gives the line profile (squared modulus of amplitude) of the corresponding process. 

We start from the S-matrix element describing the scattering of two photons on the atomic electron with consequent emission of one photon \cite{ZSLP-report}
\begin{eqnarray}
\label{1}
\hat{S}^{(3)}_{if}=(-\mathrm{i}e)^3\int d^4x_{3}d^4x_{2}d^4x_1
\overline{\psi}_{f}(x_{3})\gamma_{\mu_{3}}A^*_{\mu_{3}}(x_3)
\\\nonumber
\times
S(x_3,x_2)\gamma_{\mu_2}A_{\mu_{2}}(x_2)S(x_2,x_1)\gamma_{\mu_1}A_{\mu_{1}}(x_1)\psi_{i}
(x_1)
.
\end{eqnarray} 
Here $\psi_a(x) = \psi_a(\textbf{r})e^{-i E_a t}$, $\psi_a(\textbf{r})$ is the solution of  Dirac equation for the atomic electron, $E_a$ is the Dirac energy, $\overline{\psi}_{a} = \psi_{a}^{+} \gamma_0$ is the Dirac conjugated wave function with $\psi_{a}^{+}$ being its Hermitian conjugate, $\gamma_{\mu} = (\gamma_0, \bm{\gamma})$ are the Dirac matrices, $S(x_1,x_2)$ is the electron propagator, $ A_{\mu}(x) $ is the photon wave function
\begin{eqnarray}
A_{\mu}(x)=\sqrt{\frac{2\pi}{\omega}}e_{\mu}^{(\lambda)}e^{\mathrm{i}k_{\mu}x_{\mu}}
\end{eqnarray}
where $k_{\mu}$ is the photon momentum 4-vector, $\textbf{k}$ is the photon wave vector, $\omega = |\textbf{k}|$ is the photon frequency, $e^{(\lambda)}_{\mu}$ are the components of the photon polarization 4-vector ($ \mu,\;\lambda = 0,1,2,3 $), $ x^{\mu}$ is the space 4-vector, $A_{\mu}$ corresponds to the absorbed photon and $A_{\mu}^{*}$ represents the emitted photon, respectively. The eigenmode decomposition of $S(x_1,x_2)$ in Eq. (\ref{1}) with respect to one-electron eigenstates is
\begin{eqnarray}
\label{2}
S(x_1,x_2) = \frac{\mathrm{i}}{2\pi}\int\limits_{-\infty}^\infty d\Omega\, e^{-\mathrm{i}\Omega(t_1-t_2)}
%\\\nonumber
%\times
\sum\limits_n\frac{\psi_n(\textbf{r}_1)
\overline{\psi}_n(\text{r}_2)}{\Omega-E_n(1-\mathrm{i}0)},\qquad
\end{eqnarray}
where summation runs over the entire Dirac spectrum.

Describing real photons implies the transversality condition
\begin{eqnarray}
\label{3}
\gamma_{\mu}e^{(\lambda)}_{\mu}=   \textbf{e}\;\bm{\alpha},
\end{eqnarray}
where $  \textbf{e} $ is the transverse 3-vector of the photon polarization. Then real transverse photons are described by
\begin{eqnarray}
\label{4}
\textbf{A}^{(\textbf{k},\textbf{e})}(x)=\sqrt{\frac{2\pi}{\omega}}\textbf{e}\;e^{\mathrm{i}(\textbf{k}\textbf{r}-\omega t)}\equiv\textbf{A}_{\textbf{k},\textbf{e}}(\textbf{r})\;e^{-\mathrm{i}\omega t},
\end{eqnarray} 
together with notation
\begin{eqnarray}
\label{5}
\textbf{A}_{\textbf{k}, \textbf{e}} = \sqrt{\frac{2\pi}{\omega}} \textbf{e}\;e^{\mathrm{i}\textbf{k}\textbf{r}}
.
\end{eqnarray} 
Insertion of the expressions (\ref{2}), (\ref{4}) into Eq. (\ref{1}) and performing the integrations over time variables and frequencies in electron propagators in S-matrix Eq. (\ref{1}) yields
\begin{eqnarray}
\label{6}
\hat{S}_{if}^{(3)}=-2\pi \mathrm{i}\;\delta(E_{f}-E_{i}+\omega_1-\omega_2-\omega_3)U^{(3)}_{if},
\end{eqnarray}
where
\begin{eqnarray}
\label{7}
U^{(3)}_{if}=e^3
\sum\limits_{nk}
\frac{
\langle  f      |
\bm{\alpha}\textbf{A}^*_{\textbf{k}_1,\textbf{e}_1}
|       n       \rangle
\langle n       |
\bm{\alpha}\textbf{A}_{\textbf{k}_2,\textbf{e}_2}
|       k       \rangle
}
{(E_{n} - E_{f} - \omega_1)}
\\\nonumber
\times\frac{\langle k       |
\bm{\alpha}\textbf{A}_{\textbf{k}_3,\textbf{e}_3}
|       i       \rangle}{ (E_k - E_{i} - \omega_2)}
+(5\;\mbox{permutations}).
\end{eqnarray}
Permutations in Eq. (\ref{7}) are understood as all possible rearrangement of indices $1,2,3$ denoting the corresponding photons. 
Then the differential cross section of the scattering process is 
\begin{eqnarray}
\label{8}
\sigma_{if}(\omega)=2\pi\delta(E_{f}-E_{i}+\omega_1-\omega_2-\omega_3)
%\\\nonumber\times
\left|U^{(3)}_{if}\right|^2
\\\nonumber
\times
d\omega_1
d\omega_2
d\omega_3
.
\end{eqnarray}

In the nonrelativistic limit Eq. (\ref{7}) takes the form \cite{Andr}
\begin{widetext}
\begin{eqnarray}
\label{9}
U^{(3)}_{if}=e^3\frac{(2\pi)^{3/2}}{\sqrt{\omega_{1}\omega_{2}\omega_{2}}}
\sum\limits_{nk}\frac{
\langle f |
\textbf{e}_1^*\textbf{p}
| n  \rangle
\langle  n |
\textbf{e}_2\textbf{p}
| k \rangle
\langle k |
\textbf{e}_3\textbf{p}
| i \rangle
}
{(E_{n} - E_{f} - \omega_1) (E_k - E_{i} - \omega_2)}
+(5\;\mbox{permutations}),
\end{eqnarray}
\end{widetext}
where $\textbf{p}$ is the electron momentum operator.

We are interested in the case when two incident photons are absorbed in some intermediate state $ n $, i.e. $ \omega_2+\omega_3=E_{n}-E_{i} $.  
In the resonant approximation, nonresonant terms in scattering amplitude can be omitted. Such an approximation is justified by the fact that the corresponding nonresonant corrections go beyond the accuracy of experiments \cite{S-2020-importance}. Then assuming that frequencies of two incident laser photons are equal, i.e.
\begin{eqnarray}
\label{equal}
\omega_2=\omega_3\equiv\omega
, 
\end{eqnarray}
the cross section Eq. (\ref{8}) with nonrelativistic scattering amplitude Eq. (\ref{9}) can be written in the form
\begin{eqnarray}
\label{10}
\sigma(\omega)= e^6
\left|
\sum\limits_{nk}
(E_n-E_f)^{3/2}\left(E_n-E_i-\omega \right)^{3/2}
\right.
\\\nonumber
\times
\omega^{3/2}
\frac{\langle f | \textbf{e}_{1}^*\textbf{r} | n \rangle}{E_{n}-E_{i}-2\omega-\frac{\mathrm{i}\Gamma_{n}}{2}}
\left\lbrace
\frac{
	\langle n | \textbf{e}_{2}\textbf{r} | k \rangle
	\langle k | \textbf{e}_{3}\textbf{r}  | i \rangle
}
{
	E_{k}-E_{i}-\omega
}
\right.
\\\nonumber
\left.
\left.
+
\frac{
	\langle n | \textbf{e}_{3}\textbf{r}| k \rangle
	\langle k | \textbf{e}_{2}\textbf{r}| i \rangle
}
{
	E_{k}-E_{n}+\omega
}
\right\rbrace
\right|^2
d\omega
,
\end{eqnarray}
where regularization procedure of divergent denominator and relation $ \langle a | \textbf{p} | b \rangle = \mathrm{i}(E_{a}-E_{b})\langle a | \textbf{r} | b \rangle$ were applied \cite{Andr}. The appearance of the imaginary part leads to the formation of the absorption line profile \cite{Andr}. 

To introduce the NR correction to the cross section given by Eq. (\ref{10}) we have to take into account all neighboring terms in the sum over $ n $. Then in case of only two neighboring states, only terms in Eq. (\ref{10}) with $ n=a $ (leading resonant term to which we introduce NR correction) and $ n=b$ (closest by energy to the resonant term) need to be considered \cite{LSPS,can-LSPS,PRA-LSPS,PRL-LSSP,LSSCK-2009}. The set of quantum numbers for the latter additional state $ b $ should allow the connection of this state with the initial state by absorption of two-photon in two electric dipole transitions. In what follows we will consider the NR corrections originating from the closest states permitted by two-photon selection rules \cite{grynbergPHD}. The neighboring resonant level may give a noticeable NR correction as it was recently observed in one-photon case \cite{beyer}. We will also neglect the contribution quadratic in NR correction and will neglect the level width in the energy denominator corresponding to the NR state. 

Then keeping in the sum over $ n $ in Eq. (\ref{11}) only neighbouring states $ a $ and $ b $ separated by the interval $ \Delta = E_{n_al_aj_aF_a}-E_{n_bl_bj_bF_b} $, performing summation over projections of final state and averaging over projections of initial state we arrive at
\begin{widetext}
\begin{eqnarray}
\label{11}
\sigma_{if}(\omega)=\frac{e^6}{2F_{i}+1}\sum\limits_{M_{F_i}M_{F_f}}
\left[
\frac{|T_{fai}(\omega)|^2}{(\omega_{ai}-2\omega)^2+\frac{\Gamma_{a}^2}{4}}+2\mathrm{Re}\frac{T_{fai}(\omega)T_{fbi}^*(\omega)}{(\omega_{ai}-2\omega-\frac{\mathrm{i}\Gamma_{a}}{2})(\omega_{bi}-2\omega)}
\right]
d\omega
,
\end{eqnarray}
where
\begin{eqnarray}
\label{12}
T_{fni}(\omega)=\omega_{nf}^{3/2}\omega^{3/2}\left(\omega_{ni}-\omega \right)^{3/2}
\sum\limits_{M_{F_n}}
\langle f | \textbf{e}_{1}^*\textbf{r} | n \rangle
\sum\limits_{kl_kj_kF_kM_{F_k}}
\left\lbrace
\frac{
	\langle n | \textbf{e}_{2}\textbf{r} | k \rangle
	\langle k | \textbf{e}_{3}\textbf{r}  | i \rangle
}
{
	\omega_{ki}-\omega
}
+
\frac{
	\langle n | \textbf{e}_{3}\textbf{r}| k \rangle
	\langle k | \textbf{e}_{2}\textbf{r}| i \rangle
}
{
	\omega_{kn}+\omega
}
\right\rbrace
,
\end{eqnarray}
%\begin{eqnarray}
%\label{13}
%T_{b}=
%\omega_{bf}^{3/2}\omega^{3/2}\left(\omega_{bi}-\omega \right)^{3/2}
%\sum\limits_{M_{F_b}}
%\frac{\langle f | \textbf{e}_{1}^*\textbf{r} | b \rangle}{\omega_{ai}+\Delta-2\omega-\frac{\mathrm{i}\Gamma_{b}}{2}}
%\sum\limits_{kl_kj_kF_kM_{F_k}}
%\left\lbrace
%\frac{
%	\langle b | \textbf{e}_{2}\textbf{r} | k \rangle
%	\langle k | \textbf{e}_{3}\textbf{r}  | i \rangle
%}
%{
%	\omega_{ki}-\omega
%}
%+
%\frac{
%	\langle b | \textbf{e}_{3}\textbf{r}| k \rangle
%	\langle k | \textbf{e}_{2}\textbf{r}| i \rangle
%}
%{
%	\omega_{kb}+\omega
%}
%\right\rbrace
%,
%\end{eqnarray}
\end{widetext}
and $ \omega_{n'n}\equiv E_{n'l'j'F'}-E_{nljF} $. The first term in Eq. (\ref{11}) corresponds to the two-photon resonant scattering on atomic state $ a $, while the last term describes the leading interference contribution to the cross section. 

In the nonrelativistic limit the matrix elements in Eq. (\ref{12}) do not depend explicitly on the photon directions $ \bm{\nu}_1 $, $ \bm{\nu}_2 $ and $ \bm{\nu}_3 $. Implicitly this dependence enters via the transversality condition. Without loss of generality one can assume that in the experiment incident photons are propagating in opposite directions with vectors of polarizations $ \textbf{e}_{2} $ and  $ \textbf{e}_{3} $, while outgoing photon has a polarization $ \textbf{e}_1 $. Following \cite{S-2020-importance}, the analysis of angular correlations of Eq. (\ref{11}) can be performed. It is supposed that in the experiment the polarizations of incident photons (propagating in opposite directions) $ \textbf{e}_2 $, $ \textbf{e}_3 $ and outgoing photon direction $ \bm{\nu}_1 $ are fixed. Then denoting the angles between any pair of two vectors as $ \theta_{ij} $ ($ i,\,j=1,\,2,\,3 $) we are interested in dependence of interference contribution in Eq. (\ref{11}) on the corresponding angles. This is exactly the situation similar to experiments based on one- and two-photon scattering.  

The dependence of cross section Eq. (\ref{11}) on $ \bm{\nu}_1 $ becomes explicit after summation over photon polarization $ \textbf{e}_{1} $. Then to describe the  experiment we have to evaluate $ \sum\limits_{\textbf{e}_1}\sigma_{if} $. It should be noted that in experiments with two-photon excitation relative orientation of polarization of incident photons is chosen to maximize the absorption rate \cite{Budker}. This is usually achieved when vectors $ \textbf{e}_2 $ and $ \textbf{e}_3 $ are parallel ($ \theta_{23}=0 $) or orthogonal to each other ($ \theta_{23}=\pi/2 $).

For evaluating the angular correlations in cross section with Eq. (\ref{11}) we employ the techniques of irreducible tensor operators (we follow notations given in \cite{S-2020-importance}). Assuming that $ \omega_{bi}=\omega_{ai}+\Delta $ the scattering cross section can be written in the form 
\begin{gather}
\label{sigmamaintext}
\sigma_{if}(\omega)=\frac{e^6}{2F_{i}+1}
%\left(
%\sum\limits_{k}\frac{I_{fa}I_{ak}I_{ki}}{E_{k}+\frac{\omega_{ai}}{2}}
%\right)^2
%\\\nonumber
%\times
\left[
\frac{f^a_{\mathrm{res}}}{(\omega_{ai}-2\omega)^2+\frac{\Gamma_{a}^2}{4}}
%\right.
%\\\nonumber
%\left.
%+
%\frac{f_{\mathrm{res}}\;b}{(\omega_{ai}-2\omega +\Delta)^2+\frac{\Gamma_{b}^2}{4}}
\right.
\\\nonumber
\left.
+2\mathrm{Re}\frac{f^{ab}_{\mathrm{nr}}}{(\omega_{ai}-2\omega-\frac{\mathrm{i}\Gamma_{a}}{2})(\omega_{ai}-2\omega+\Delta )}
%\\\nonumber
%\times
%\left.
%\frac{1}{\omega_{ai}-2\omega+\Delta + \frac{\mathrm{i}\Gamma_{b}}{2}}
\right]d\omega
,
\end{gather}
where 
\begin{eqnarray}
\label{fres}
f^{a}_{\mathrm{res}}=\sum\limits_{M_{F_i}M_{F_f}}|T_{fai}(\omega_{ai}/2)|^2
\end{eqnarray}
 and 
\begin{eqnarray}
\label{fnr}
f^{ab}_{\mathrm{nr}} =\sum\limits_{M_{F_i}M_{F_f}}T_{fai}(\omega_{ai}/2)T^*_{fbi}(\omega_{ai}/2)
\end{eqnarray}
are the coefficients defining angular dependencies. Their analytical evaluation is presented in Appendix A.

The resonant transition frequency $ \omega_{\mathrm{res}} $ can be defined from $ \sigma_{if}(\omega) $ by different ways. One evident way is to define $ \omega_{\mathrm{res}} $ as $ \omega_{\mathrm{res}}=\omega_{\mathrm{max}} $, where $ \omega_{\mathrm{max}} $ corresponds to the maximum value of $ \sigma_{if}(\omega) $ . Then $ \omega_{\mathrm{res}} $ can be obtained from the condition  
\begin{eqnarray}
\label{15}
\frac{d}{d\omega}\sigma_{if}(\omega)=0
.
\end{eqnarray}
In the resonant approximation (i.e. retaining only first term in Eq. (\ref{sigmamaintext})) we immediately find $ \omega_{\mathrm{res}}=\omega_{\mathrm{max}}=\omega_{ai}=E_{n_al_aj_aF_a}-E_{n_il_ij_iF_i} $. However, keeping the last interference term in Eq. (\ref{sigmamaintext}) and solving Eq. (\ref{15}) with respect to $  \omega$ we arrive at the definition $ \omega_{\mathrm{max}} $
\begin{eqnarray}
\label{omegamax}
\omega_{\mathrm{max}}=\omega_{ai}-\delta_{\mathrm{NR}}
,
\end{eqnarray}
where
\begin{eqnarray}
\label{nr}
\delta_{\mathrm{NR}}=\frac{f^{ab}_{\mathrm{nr}}}{f^a_{\mathrm{res}}}\frac{\Gamma_{a}^2}{4\Delta}
.
\end{eqnarray} 
NR correction Eq. (\ref{nr}) is obtained as the lowest term of an expansion of the result in terms of $ \Gamma/\Delta $ \cite{S-2020-importance}, when this parameter is small. The angular dependence of NR correction Eq. (\ref{nr}) is defined by the ratio of $ f^{ab}_{\mathrm{nr}}/f^a_{\mathrm{res}}$  and depends on the arrangement of the experiment, i.e. on the angles between each pair of vectors $ \bm{\nu}_1 $, $\textbf{e}_2 $ and  $\textbf{e}_3 $.

\section{Application to $ 2s_{1/2}^{F=0}\rightarrow ns_{1/2}^{F=0} $ and $ 2s_{1/2}^{F=1}\rightarrow ns_{1/2}^{F=1} $ transitions}
\label{section2}

In this section we consider NR correction orgiginating from neighbouring $ nd $ states for two-photon transitions $ 2s_{1/2}^{F=0}\rightarrow ns_{1/2}^{F=0} $ and $ 2s_{1/2}^{F=1}\rightarrow ns_{1/2}^{F=1} $ (with $ n=4,\,6,\,8,\,12 $) assuming that hyperfine structure of $ 2s $ electron shell is resolvable in experiments \cite{deB-0,deB-1,Schwob,Weitz-1,Nez,beyer}. According to the two-photon selection rules the spin-flip electric dipole two-photon transitions $ 2s_{1/2}^{F=0}\rightarrow ns_{1/2}^{F=1} $ or  $ 2s_{1/2}^{F=1}\rightarrow ns_{1/2}^{F=0} $ are strongly suppressed \cite{SSSR-2015,SSSR-2016,SSSR-2017}. Therefore, the interference with close $ nd $ states is only possible. 
As a result $ 2s_{1/2}^{F=1}\rightarrow ns_{1/2}^{F=1} $ transition branch interfere with $ 2s_{1/2}^{F=1}\rightarrow nd_{3/2}^{F=1} $, $ 2s_{1/2}^{F=1}\rightarrow nd_{3/2}^{F=2} $, $ 2s_{1/2}^{F=1}\rightarrow nd_{5/2}^{F=2} $ and $ 2s_{1/2}^{F=1}\rightarrow nd_{5/2}^{F=3} $ two-photon absorption branches \cite{science2020}. Then for $ 2s_{1/2}^{F=1}\rightarrow ns_{1/2}^{F=1} $
we set in all equations $ n_{i} l_{i}=2s $, $ j_i=1/2 $, $ F_{i}=1 $, $ n_{a} l_{a}=ns $, $ n_{b} l_{b}=nd $ (with $ n_{a}=n_{b}=4,\,6,\,8,\,12 $), $ j_{a}=1/2 $, $ F_{a}=1 $. Performing summation over all possible values of $ j_{b} $ and $ F_{b} $ the leading total NR correction can be written as a sum of four contributions 
\begin{gather}
\label{nrsum2}
\delta_{\mathrm{NR}}(2s_{1/2}^{F=1}-ns_{1/2}^{F=1} )=
\frac{\Gamma_{ns_{1/2}}^2}{4\sum\limits_{j_{f}F_{f}}f_{\mathrm{res}}^{ns_{1/2}^{F=1}}}
\\\nonumber
\times
\left(
\frac{\sum\limits_{j_{f}F_{f}}f_{\mathrm{nr}}^{ns_{1/2}^{F=1} nd_{3/2}^{F=1}}}{\Delta_{1}}
+
\frac{\sum\limits_{j_{f}F_{f}}f_{\mathrm{nr}}^{ns_{1/2}^{F=1} nd_{3/2}^{F=2}}}{\Delta_{2}}
\right.
\\\nonumber
\left.
+
\frac{\sum\limits_{j_{f}F_{f}}f_{\mathrm{nr}}^{ns_{1/2}^{F=1} nd_{5/2}^{F=2}}}{\Delta_{3}}
+
\frac{\sum\limits_{j_{f}F_{f}}f_{\mathrm{nr}}^{ns_{1/2}^{F=1} nd_{5/2}^{F=3}}}{\Delta_{4}}
\right)
,
\end{gather} 
where $ \Delta_1=E_{ns_{1/2}^{F=1}}-E_{nd_{3/2}^{F=1}}$, $ \Delta_2=E_{ns_{1/2}^{F=1}}-E_{nd_{3/2}^{F=2}} $, $ \Delta_3=E_{ns_{1/2}^{F=1}}-E_{nd_{5/2}^{F=2}} $,  $ \Delta_4=E_{ns_{1/2}^{F=1}}-E_{nd_{5/2}^{F=3}} $ and $ \Gamma_{ns_{1/2}}$ is the level width of $ ns_{1/2} $ state (here and below we assume that $ \Gamma_{nlj}=\Gamma_{nljF} $). 

\begin{figure}[hbtp]
\centering
\caption{Laser frequency shifts $ \delta_{\mathrm{NR}}/2 $ (in Hz) for the measurement of $2s_{1/2}^{F=1}-ns_{1/2}^{F=1}$ ($ n=4,\,6,\,8,\,12 $) transition frequencies in hydrogen, see Eq. (\ref{f4})-(\ref{f12}). The horizontal axes is the
angle  $ \theta $  between polarization vector $ \textbf{e}_2 $ of incident photon (or $ \textbf{e}_3 $, since $ \textbf{e}_2\parallel\textbf{e}_3 $) and propagation direction $ \bm{\nu}_1 $ of outgoing photon. All shifts turn to zero at 'magic' angles $ \theta=54.7^{\circ}  $ and $ \theta=125.3^{\circ} $.
}
\includegraphics[scale=0.5]{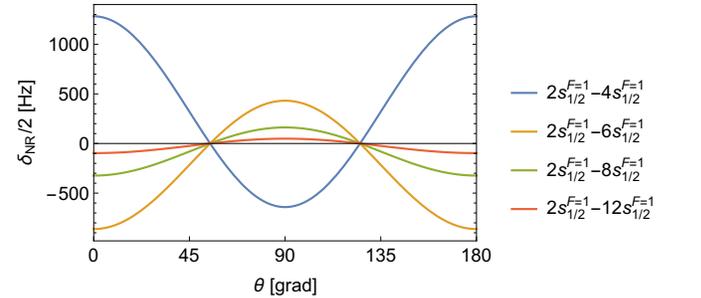}
\label{fig1}
\end{figure}

In the experiments \cite{crossdamping,yost} the polarization of incident laser photons $ \textbf{e}_2 $ and $ \textbf{e}_3 $ were fixed as parallel to each other. Then the NR correction, Eq. (\ref{nrsum2}), depends only on one angle between polarization of outgoing photon $ \textbf{e}_1 $ and one of the two parallel vectors $ \textbf{e}_2 $ or $ \textbf{e}_3 $. Summation over the polarization $ \textbf{e}_1 $ leads to an implicit dependence on the vector of propagation direction $ \bm{\nu}_1 $. Denoting the angle between vectors $ \textbf{e}_2 $ (or $ \textbf{e}_3 $) and $ \bm{\nu}_1 $  as $ \theta $, substituting all numerical values of levels widths, energy differences into Eq. (\ref{nrsum2}) and evaluating the sum over entire spectrum in Eq. (\ref{12}) (see details in Appendix B) we arrive at the following numerical results for NR corrections (in Hz)

\begin{eqnarray}
\label{f4}
\delta_{\mathrm{NR}}(2s_{1/2}^{F=1}-4s_{1/2}^{F=1} )=640.80(1+3\cos(\theta))
,
\end{eqnarray}
\begin{eqnarray}
\label{f6}
\delta_{\mathrm{NR}}(2s_{1/2}^{F=1}-6s_{1/2}^{F=1} )=-431.12(1+3\cos(\theta))
,
\end{eqnarray}
\begin{eqnarray}
\label{f8}
\delta_{\mathrm{NR}}(2s_{1/2}^{F=1}-8s_{1/2}^{F=1} )=-161.67(1+3\cos(\theta))
,
\end{eqnarray}
\begin{eqnarray}
\label{f12}
\delta_{\mathrm{NR}}(2s_{1/2}^{F=1}-12s_{1/2}^{F=1} )=-48.51(1+3\cos(\theta))
.
\end{eqnarray}
The angular correlation of NR corrections, Eqs. (\ref{f4})-(\ref{f12}), are illustrated in Fig. \ref{fig1}. It implicitly shows the absence of interference at 'magic' angles $ \theta=54.7^{\circ}  $ and $ \theta=125.3^{\circ} $. In general case with arbitrary arrangement of vectors $\textbf{e}_2$, $\textbf{e}_3$ and $ \bm{\nu}_1 $ NR correction is given by Eq. (\ref{nrsum3}) in Appendix B.

For numerical evaluation we used theoretical values for energies given in \cite{HH-tab}, which incorporate relativistic, QED, nuclear size, the hyperfine structure corrections. The same concerns the value of the widths and intervals $ \Delta $. These values give a sufficiently accurate result for $ \delta_{\mathrm{NR}} $. The parameter $ \Gamma /\Delta $ is always $ \ll 1 $ for all considered cases, which makes the used approximation adequate.

Recently the similar interference effects in $ 1s-3s $ spectroscopy of hydrogen were studied in \cite{crossdamping}. It was found that for two-photon laser induced transition $ 1s_{1/2}^{F=1}\rightarrow 3s_{1/2}^{F=1} $ the NR correction due to the interference with four neighbouring transitions $ 1s_{1/2}^{F=1}\rightarrow 3d_{3/2}^{F=1} $, $ 1s_{1/2}^{F=1}\rightarrow 3d_{3/2}^{F=2} $, $ 1s_{1/2}^{F=1}\rightarrow 3d_{5/2}^{F=2} $ and $ 1s_{1/2}^{F=1}\rightarrow 3d_{5/2}^{F=3} $ is less than experimental uncertainty. Equation (\ref{nrsum2}) can be easily extended to the calculation of NR correction to $ 1s_{1/2}^{F=1}\rightarrow 3s_{1/2}^{F=1} $ transition frequency by replacing $ 2s_{1/2}^{F=1}\leftrightarrow 1s_{1/2}^{F=1} $ and setting $ n_{a}=n_{b}=3 $. 
Then the NR correction is (in Hz)
\begin{eqnarray}
\label{nr1s3s}
\delta_{\mathrm{NR}}(1s_{1/2}^{F=1}-3s_{1/2}^{F=1} )=-225.61(1+3\cos(\theta))
.
\end{eqnarray}

\begin{figure}[hbtp]
\centering
\caption{Laser frequency shift $ \delta_{\mathrm{NR}}/2 $ (in Hz) for the measurement of $1s_{1/2}^{F=1}-3s_{1/2}^{F=1}$ 
 transition frequency in hydrogen, see Eq. (\ref{nr1s3s}). All notations are the same as in Fig. \ref{fig1}.}
\includegraphics[scale=0.5]{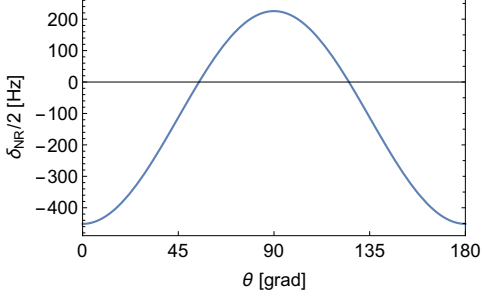}
\label{fig2}
\end{figure}

According to Eq. (\ref{nr1s3s}) the corresponding laser frequency shift which could be defined as $ \delta^{(2)}_{\mathrm{NR}}/2 $ reaches its maximum value $ -0.446 $ kHz at $ \theta=0 $, see Fig. \ref{fig2}. This is in excellent agreement with the previous result $ -0.45 $ kHz obtained in a completely different way in \cite{crossdamping,yost} (see Figs. 4 and 5 in \cite{crossdamping} and \cite{yost}, respectively), whereas the experiental uncertainty in the recent measurement of $ 1s-3s $ transition frequency is $ 0.72 $ kHz \cite{science2020}.

For other transition, $ 2s_{1/2}^{F=0}\rightarrow ns_{1/2}^{F=0} $, there is interference with $ 2s_{1/2}^{F=0}\rightarrow nd_{3/2}^{F=2} $ and $ 2s_{1/2}^{F=0}\rightarrow nd_{5/2}^{F=2} $ two-photon absorption branches. In this case the leading total NR correction can be written as a sum of two contributions 
\begin{gather}
\label{nrsum4}
\delta_{\mathrm{NR}}(2s_{1/2}^{F=0}-ns_{1/2}^{F=0} )=
\frac{\Gamma_{ns_{1/2}}^2}{4\sum\limits_{j_{f}F_{f}}f_{\mathrm{res}}^{ns_{1/2}^{F=0}}}
\\\nonumber
\times
\left(
\frac{\sum\limits_{j_{f}F_{f}}f_{\mathrm{nr}}^{ns_{1/2}^{F=0}nd_{3/2}^{F=2}}}{\Delta_{1}'}
+
\frac{\sum\limits_{j_{f}F_{f}}f_{\mathrm{nr}}^{ns_{1/2}^{F=0}nd_{5/2}^{F=2}}}{\Delta_{2}'}
\right)
,
\end{gather}
where now $ \Delta_1'=E_{ns_{1/2}^{F=0}}-E_{nd_{3/2}^{F=2}} $ and $ \Delta_2'=E_{ns_{1/2}^{F=0}}-E_{nd_{5/2}^{F=2}} $. 

\begin{figure}[hbtp]
\centering
\caption{Laser frequency shift $ \delta_{\mathrm{NR}}/2 $ (in Hz) for the measurement of $2s_{1/2}^{F=0}-ns_{1/2}^{F=0}$ ($ n=4,\,6,\,8,\,12 $) transition frequencies in hydrogen, see Eq. (\ref{g4})-(\ref{g12}).  All notations are the same as in Fig. \ref{fig1}.}
\includegraphics[scale=0.5]{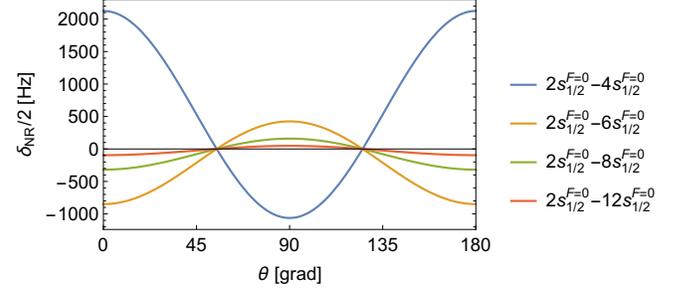}
\label{fig3}
\end{figure}

Evaluation of Eq. (\ref{nrsum4}) for the case when atom in its metastable  $ 2s_{1/2}^{F=0} $ state is excited by two equal frequency laser photons with parallel polarizations $ \textbf{e}_2 $ and $ \textbf{e}_3 $ is presented in Appendix B. Then the final results for NR corrections are (in Hz)
\begin{eqnarray}
\label{g4}
\delta^{(2)}_{\mathrm{NR}}(2s_{1/2}^{F=0}-4s_{1/2}^{F=0} )=1062.57(1+3\cos(\theta))
,
\end{eqnarray}
\begin{eqnarray}
\label{g6}
\delta_{\mathrm{NR}}(2s_{1/2}^{F=0}-6s_{1/2}^{F=0} )=-424.51(1+3\cos(\theta))
\end{eqnarray}
\begin{eqnarray}
\label{g8}
\delta_{\mathrm{NR}}(2s_{1/2}^{F=0}-8s_{1/2}^{F=0} )=-159.19(1+3\cos(\theta))
\end{eqnarray}
\begin{eqnarray}
\label{g12}
\delta_{\mathrm{NR}}(2s_{1/2}^{F=0}-12s_{1/2}^{F=0} )=-47.76(1+3\cos(\theta))
.
\end{eqnarray}
The dependence on $ \theta $ is illustrated in Fig. \ref{fig3} where it can be found again that the 'magic angles' give the vanishing of NR corrections Eqs. (\ref{g4})-(\ref{g12}). In general case with arbitrary arrangement of vectors $\textbf{e}_2$, $\textbf{e}_2$ and $ \bm{\nu}_1 $ NR correction is given by Eq. (\ref{nrsum5}) in Appendix B.

\section{Application to $ 2s_{1/2}^{F=1}\rightarrow nd_{3/2}^{F=2} $ and $ 2s_{1/2}^{F=1}\rightarrow nd_{5/2}^{F=2} $ transitions}
\label{section3}

Now we turn to evaluation of NR correction to $ 2s_{1/2}^{F=1}\rightarrow nd_{3/2}^{F=2} $ ($ n=4,\,6,\,8,\,12 $) transition frequencies with the account for neighbouring $ nd_{5/2}^{F=2} $ level. For this purpose we set in all equations $ n_{i} l_{i}=2s $, $ j_i=1/2 $, $ F_{i}=1 $, $ n_{a} l_{a}=n_{b} l_{b}=nd $ (with $ n_{a}=4,\,6,\,8,\,12 $), $ j_{a}=3/2 $, $ F_{a}=2 $, $ j_{b}=3/2 $, $ F_{b}=2 $.  Considering the case of experiment \cite{Weitz} when the fine and hyperfine sublevels of the final state are not resolvable by the detector, the summation over quantum numbers $ j_{f}F_{f} $ in Eq. (\ref{sigmamaintext}) should be performed. This leads to the NR correction
\begin{eqnarray}
\label{nrsumnd1}
\delta_{\mathrm{NR}}(2s_{1/2}^{F=1}-nd_{3/2}^{F=2})=\frac{\sum\limits_{j_{f}F_{f}}f_{\mathrm{nr}}^{nd_{3/2}^{F=2}nd_{5/2}^{F=2}}}{\sum\limits_{j_{f}F_{f}}f_{\mathrm{res}}^{nd_{3/2}^{F=2}}}
\frac{\Gamma_{nd_{3/2}}^2}{4\Delta''}
,
\end{eqnarray} 
where $ \Delta''=E_{nd_{3/2}^{F=2}}-E_{nd_{5/2}^{F=2}} $. The numerical evaluation of Eq. (\ref{nrsumnd1}) is similar to previous cases considered in section \ref{section2}, see Eq. (\ref{nrsumnd2}) in Appendix B.  Finally, we arrive at the NR corrections which do not depend on the angles between the  vectors $ \bm{\nu}_1 $, $ \textbf{e}_2 $ and $ \textbf{e}_3 $ (in Hz) 
\begin{eqnarray}
\label{e4}
\delta_{\mathrm{NR}}(2s_{1/2}^{F=1}-4d_{3/2}^{F=2} )=967.75
\end{eqnarray}
\begin{eqnarray}
\label{e6}
\delta_{\mathrm{NR}}(2s_{1/2}^{F=1}-6d_{3/2}^{F=2} )=296.48
\end{eqnarray}
\begin{eqnarray}
\label{e8}
\delta_{\mathrm{NR}}(2s_{1/2}^{F=1}-8d_{3/2}^{F=2} )=127.31
\end{eqnarray}
\begin{eqnarray}
\label{e12}
\delta_{\mathrm{NR}}(2s_{1/2}^{F=1}-12d_{3/2}^{F=2} )=38.38
.
\end{eqnarray}

It is important to note that numerical values of NR corrections given by Eqs. (\ref{e4})-(\ref{e12}) are turned out several times less than the same NR corrections calculated recently in \cite{solovyev2020proton} for the other type of experiment monitoring the rate of $ 2s $ state quenching \cite{deB-1,deB-2}. The difference with present results actually arises due to the rejected transiton matrix element of the emitted photon in the amplitude of scattering process used in \cite{solovyev2020proton}. It can be concluded that the detection of the excited $ ns/nd$ fraction of atoms via its fluorescence has a considerably higher potential accuracy than the experiments monitoring the rate of metastable state quenching \cite{Weitz}. 

\begin{figure}[hbtp]
\centering
\caption{Total laser frequency shift $ \delta_{\mathrm{NR}}/2 $ (in Hz) for the measurement of $2s_{1/2}^{F=1}-nd_{3/2}^{F=2}$ ($ n=4,\,6,\,8,\,12 $) transition frequencies in hydrogen, see Eqs. (\ref{e4})-(\ref{e12}) and Eqs. (\ref{h4})-(\ref{h12}).  All notations are the same as in Fig. \ref{fig1}.}
\includegraphics[scale=0.5]{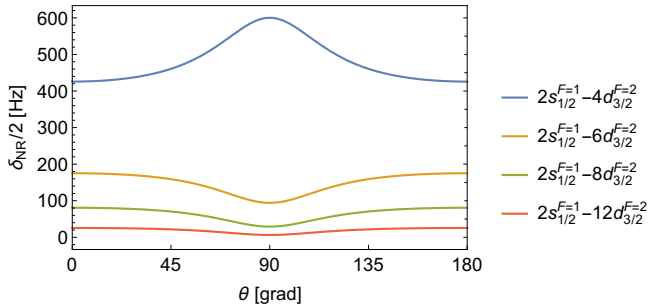}
\label{fig4}
\end{figure}

Besides the corrections Eqs. (\ref{e4})-(\ref{e12}) arising due to the neighbouring $ nd_{3/2}^{F=2} $ and $ nd_{5/2}^{F=2} $ states, the quantum interference between $ 2s_{1/2}^{F=1}\rightarrow nd_{3/2}^{F=2} $ and $ 2s_{1/2}^{F=1}-ns_{1/2}^{F=1}$ absorption branches should be considered. Similar to Eq. (\ref{nrsumnd1}) we can write
\begin{eqnarray}
\label{nrsumnd22}
\delta_{\mathrm{NR}}(2s_{1/2}^{F=1}-nd_{3/2}^{F=2})=\frac{\sum\limits_{j_{f}F_{f}}f_{\mathrm{nr}}^{nd_{3/2}^{F=2}ns_{1/2}^{F=1}}}{\sum\limits_{j_{f}F_{f}}f_{\mathrm{res}}^{nd_{3/2}^{F=2}}}
\frac{\Gamma_{nd_{3/2}}^2}{4\Delta'''}
,
\end{eqnarray} 
where $ \Delta'''= E_{nd_{3/2}^{F=2}}-E_{ns_{1/2}^{F=1}}$. Again, using Eq. (\ref{nrsumnd3}) in Appendix B, the NR corrections are (in Hz)
\begin{eqnarray}
\label{h4}
\delta_{\mathrm{NR}}(2s_{1/2}^{F=1}-4d_{3/2}^{F=1} )=
-232.602
\frac{1+3\cos(\theta)}{5+3\cos(\theta)}
,
\end{eqnarray} 
\begin{eqnarray}
\label{h6}
\delta_{\mathrm{NR}}(2s_{1/2}^{F=1}-6d_{3/2}^{F=1} )=
107.937
\frac{1+3\cos(\theta)}{5+3\cos(\theta)}
,
\end{eqnarray} 
\begin{eqnarray}
\label{h8}
\delta_{\mathrm{NR}}(2s_{1/2}^{F=1}-8d_{3/2}^{F=1} )=
68.697
\frac{1+3\cos(\theta)}{5+3\cos(\theta)}
,
\end{eqnarray} 
\begin{eqnarray}
\label{h12}
\delta_{\mathrm{NR}}(2s_{1/2}^{F=1}-12d_{3/2}^{F=1} )=
25.582
\frac{1+3\cos(\theta)}{5+3\cos(\theta)}
.
\end{eqnarray} 
The total NR correction to the  $ 2s_{1/2}^{F=1}\rightarrow nd_{3/2}^{F=2} $  transition frequencies is given by the sum of constant contributions, Eqs. (\ref{e4})-(\ref{e12}), and corresponding angular-dependent contributions, Eqs. (\ref{h4})-(\ref{h12}). The total laser frequency shifts, which are defined as $ \delta_{\mathrm{NR}}/2 $ are depicted in Fig. \ref{fig4}. It is seen that the denominator of Eqs. (\ref{e4})-(\ref{e12}) is always non-zero and positive while the numerator still turns the NR correction to zero at 'magic angles'. 

\section{Conclusion}
\label{conclusion}

In this work, the NR corrections to the $ 2s\rightarrow ns/nd $ transition frequencies in hydrogen are studied in details for experiments based on two-photon spectroscopy \cite{crossdamping,yost}. Closed analytical expressions are obtained that can be used to study any angular correlations between the polarization vectors and the directions of incident and outgoing photons. 

We restricted ourselves to studying the correlations corresponding to the experimental setup \cite{crossdamping,yost}, where an initially prepared beam of atoms in a metastable state was excited by two antiparallel laser photons with the same polarization vectors. Then, there is the only one correlation corresponding to the angle between the direction of emitted photon $ \bm{\nu}_1 $ and one of two equal vectors of incident photon polarizations $ \textbf{e}_2 $ or $ \textbf{e}_3 $. 
Other types of correlations can be easily obtained using the expressions found in this work. In addition, the results obtained in this work can be easily extended to other light atomic systems, in particular, deuterium and helium \cite{heliumtwophoton,Pachucki_1996} by setting the quantum numbers for initial, final, and intermediate states.

The analysis carried out in the work shows that for some transitions, in particular for $ 2s\rightarrow nd $, interference effects cannot be completely eliminated by choosing a certain geometry of the experiment, see Fig. \ref{fig4}. Thus, extracting the experimental parameter $\omega_{ai}$ from the measured data requires an accurate theoretical evaluation of the NR corrections. In turn, the NR corrections depend on the process used in the experiment and, therefore, should be calculated for each specific case separately, see Figs.~\ref{fig1}-{\ref{fig4}}.

The numerical values of the NR corrections can be found using expressions written in sections~\ref{section2} and \ref{section3}, specifying the correlation angle $\theta$. The magnitude of NR corrections to the transition frequencies $ 2s\rightarrow ns/nd $ decrease in a cubic manner with an increase in the principal quantum number $ n $, since the width and the corresponding splitting of energy levels are proportional to $1/n^3$. In this regard, precision measurements of the transition frequencies for Rydberg states by two-photon spectroscopy methods could also contribute to an increase in the accuracy of determining the fundamental physical constants.

\section{Acknowledgements}
This work was supported by Russian Science Foundation (Grant No. 20-72-00003).

\appendix
\renewcommand{\theequation}{A\arabic{equation}}
\setcounter{equation}{0}
\section*{Appendix A: angular algebra}

In this Appendix, the derivation of the most general expression for the photon scattering cross section on the hydrogen atom is given. In this expression, along with fine splitting, the hyperfine structure of the atomic levels is explicitly taken into account. We also present the cross sections averaged over the incident photon polarizations and summed over the outgoing photon polarizations. The closed expression for the cross section is written in terms of $ 6j $-symbols and simple radial integrals, where the angular calculations were performed using the book \cite{VMK}. 

Our goal is to obtain an analytical expression for the differential cross section with an explicit dependence on the angles between the polarization vectors of three photons. Another type of correlations including the dependence of photon propagation directions can be easily obtained from this expression by the summation over polarization.

First, we present the relations that will be useful for the subsequent derivations, see \cite{VMK}. Summation over photon polarizations can be performed with the use of the formula \cite{LabKlim}.
\begin{eqnarray}
\label{r1}
\sum_{\textbf{e}}(\textbf{e}^{*}\textbf{a})(\textbf{e}\,\textbf{b})=(\bm{\nu}\times \bm{a})(\bm{\nu}\times \bm{b}),
\end{eqnarray}
where $ \textbf{a} $, $ \textbf{b} $ are two arbitrary vectors. We denote the vector components in a cyclic basis as $ (\textbf{a})_{q} $, with $ q=0,\;\pm 1 $. In general, we will use irreducible tensors $ a_{p} $ of the rank $ p $ with the components $ a_{pq} $. The first lower index denotes the rank and the second one denotes the component. The irreducible tensor $ a_{1} $ of the rank 1 with the components $ a_{1q} $ correspond to the vector $ \textbf{a} $ and cyclic vector component $ (\textbf{a})_{q} $. The vector component $ (\textbf{a})_{q} $ is equal to the tensor component $ a_{1q} $. 

The scalar product of two arbitrary vectors $ \textbf{a} $ and $ \textbf{b} $ can be written in terms of cyclic components as
\begin{eqnarray}
\label{cyclic}
\textbf{a}\textbf{b}=\sum\limits_{q}(-1)^qa_{q}b_{-q}
.
\end{eqnarray}
The irreducible tensor product of two polarization vectors $ \textbf{e}_1 $ and $ \textbf{e}_2 $ can be expressed as follows
\begin{eqnarray}
\label{r2}
\lbrace \textbf{e}_{1}^*\otimes \textbf{e}_{2} \rbrace_{x\xi}=\sum\limits_{q_1q_2}C^{x\xi}_{1q_11q_2}(e^*_{1})_{q_{1}}(e_{2})_{q_2}
\\\nonumber
=(-1)^{\xi}\Pi_{x}
\sum\limits_{q_1q_2}
\begin{pmatrix}
   1 & 1 & x \\
   q_1 & q_2 & -\xi
\end{pmatrix}
(e^*_{1})_{q_{1}}(e_{2})_{q_2}
,
\end{eqnarray}
where $ C^{LM}_{l_1m_1l_2m_2} $ is the Clebsch-Gordan coefficient.
The irreducible tensor product $\lbrace \textbf{e}_1^*\otimes\textbf{e}_2\rbrace_{x\xi}$ satisfies the relation \cite{rappoport}:
\begin{eqnarray}
\label{r3}
\lbrace \textbf{e}_1^*\otimes\textbf{e}_2\rbrace_{x\xi}^{*}=(-1)^{x-\xi}\lbrace \textbf{e}_1\otimes\textbf{e}_2^*\rbrace_{x-\xi}
.
\end{eqnarray}
The scalar product of two irreducible tensors of rank $ x $ is
\begin{eqnarray}
\label{r4}
\sum_{\xi}a_{x\xi}b^*_{x\xi}=(a_x\cdot b_{x})
\\\nonumber
=(-1)^{-\xi}\sqrt{2x+1}\left\lbrace a_x\otimes b_{x}\right\rbrace_{00}
.
\end{eqnarray}

Permutation of three first rank tensors in mixed irreducible tensor product satisfies the following relation (see Eq. (8) on page 62 in \cite{VMK}):
\begin{eqnarray}
\label{r5}
\lbrace
\lbrace \textbf{e}_{i}\otimes \textbf{e}_{j}^* \rbrace_{x}
\otimes
\textbf{e}^*_{k}
\rbrace
_{g}
=(-1)^{x+1+g}\sum_{h}
\Pi_{xh}
\\\nonumber
\times
\begin{Bmatrix}
1 & 1 & x\\
g & 1 & h 
\end{Bmatrix}
\lbrace
\textbf{e}_{j}^*
\otimes
\lbrace \textbf{e}_{i}\otimes \textbf{e}_{k}^* \rbrace_{h}
\rbrace
_{g}
\\\nonumber
=
\sum_{h}
(-1)^{x+h}
\Pi_{xh}
\begin{Bmatrix}
1 & 1 & x\\
g & 1 & h 
\end{Bmatrix}
\\\nonumber
\times
\lbrace
\lbrace \textbf{e}_{i}\otimes \textbf{e}_{k}^* \rbrace_{h}
\otimes
\textbf{e}_{j}^*
\rbrace
_{g}
.
\end{eqnarray}
Here indices $ i,\,j,\,k=1,\,2,\,3 $ denotes the corresponding photons. Permutation of two first rank in irreducible tensor product obeys the equality
\begin{eqnarray}
\label{r5a}
\lbrace \textbf{e}_{i}\otimes \textbf{e}_{j} \rbrace_{z}=(-1)^{z}\lbrace \textbf{e}_{j}\otimes \textbf{e}_{i} \rbrace_{z}
.
\end{eqnarray}

The matrix element of the cyclic component of radius vector is given by \cite{VMK}
\begin{eqnarray}
\langle n'l'j'F'M_{F'}|r_{q}|nljFM_{F}\rangle=(-1)^{F'-M_{F'}}
\\\nonumber
\times
\begin{pmatrix}
     F' & 1 & F \\
-M_{F'} & q & M_{F}
\end{pmatrix}
\langle n'l'j'F'||r||nljF\rangle
\end{eqnarray}
where the reduced matrix element is
\begin{gather}
\label{red}
\langle n'l'j'F'||r||nljF\rangle= (-1)^{j'+j+I+l'+1/2+F}
\\\nonumber
\times
\Pi_{j'jF'F}
\begin{Bmatrix}
j' & F' & I \\
F  & j  & 1
\end{Bmatrix}
\begin{Bmatrix}
l' & j' & 1/2 \\
j  & l  & 1
\end{Bmatrix}
\langle n' l' || r|| nl \rangle
.
\end{gather}
Here $ I $ is the nuclear spin ($ I=1/2 $ for hydrogen atom) and
\begin{eqnarray}
\label{rad}
\langle n' l' || r || nl \rangle = (-1)^{l'}\Pi_{l'l}
\begin{pmatrix}
l & 1 & l'\\
0 & 0 & 0
\end{pmatrix}
\\\nonumber
\times
\int_{0}^{\infty}r^3 R_{n'l'}R_{nl}dr.
\end{eqnarray}
In Eq. (\ref{rad}) $ R_{nl} $ denotes the radial part of hydrogen wave function.

The squared nonrelativistic amplitude of three-photon scattering process is given by Eqs. (\ref{11}), (\ref{12}) in the main text.  For the evaluation of angular correlations it is convenient to consider the product $ T_{n}T_{n'}^* $, which reduces to the first and second terms of Eq. (\ref{11}) at certain values of $ n $ and $ n' $
\begin{gather}
\label{s1}
\sum\limits_{M_{F_i}M_{F_f}}T_{n}T_{n'}^*
=
\omega^3\omega_{nf}^{3/2}\omega_{n'f}^{3/2}\left(\omega_{ni}-\omega \right)^{3/2}
\left(\omega_{n'i}-\omega \right)^{3/2}
\\\nonumber
\times
\sum\limits_{\substack{M_{F_n}M_{F_{n'}}\\M_{F_k}M_{F_{k'}}\\M_{F_i}M_{F_f}}}
\sum\limits_{\substack{kl_kj_kF_k\\k'l_{k'}j_{k'}F_{k'}}}
%\\\nonumber
%&
%\times
\left\lbrace
\frac{
P_{nk}(123)
}
{
	\omega_{ki}-\omega
}
%\right.
%\\\nonumber
%&
%\left.
+
\frac{
P_{nk}(132)
}
{
	\omega_{kn}+\omega
}
\right\rbrace
\\\nonumber
\times
\left\lbrace
\frac{
P_{n'k'}(123)
}
{
	\omega_{k'i}-\omega
}
%\right.
%\\\nonumber
%&
%\left.
+
\frac{
P_{n'k'}(132)
}
{
	\omega_{k'n'}+\omega
}
\right\rbrace^*
,
\end{gather}
where 
\begin{gather}
\label{pnk}
P_{nk}(\lambda_{1}\lambda_{2}\lambda_{3})=
\langle n_{f}l_{f}j_{f}F_{f}M_{F_f} | \textbf{e}_{\lambda_{1}}^*\textbf{r} | nl_{n}j_{n}F_{n}M_{F_n} \rangle
\\\nonumber
\times
	\langle nl_{n}j_{n}F_{n}M_{F_n} | \textbf{e}_{\lambda_{2}}\textbf{r} | kl_{k}j_{k}F_{k}M_{F_k} \rangle
	\\\nonumber
\times
	\langle kl_{k}j_{k}F_{k}M_{F_k} | \textbf{e}_{\lambda_{3}}\textbf{r}  | n_{i}l_{i}j_{i}F_{i}M_{F_i} \rangle
.
\end{gather}
Eq. (\ref{s1}) reduces to the first and second terms in Eq. (\ref{11}) at $ n=n'=a $ and $ n=n'=b $, respectively. 
Then, expanding the curly braces in Eq. (\ref{s1}), we arrive at
\begin{widetext}
\begin{eqnarray}
\label{s2}
\sum\limits_{M_{F_i}M_{F_f}}T_{n}T_{n'}^*=
\omega^3\omega_{nf}^{3/2}\omega_{n'f}^{3/2}\left(\omega_{ni}-\omega \right)^{3/2}
\left(\omega_{n'i}-\omega \right)^{3/2}
\sum\limits_{\substack{M_{F_n}M_{F_{n'}}\\M_{F_k}M_{F_{k'}}\\M_{F_i}M_{F_f}}}
\sum\limits_{\substack{kl_kj_kF_k\\k'l_{k'}j_{k'}F_{k'}}}
\\\nonumber
\times
\left\lbrace
\frac{P_{nk}(123)P^*_{n'k'}(123)}
{
	(\omega_{ki}-\omega)(\omega_{k'i}-\omega)
}
+
\frac{P_{nk}(132)P^*_{n'k'}(123)}
{
	(\omega_{kn}+\omega)(\omega_{k'i}-\omega)
}
+
\frac{P_{nk}(123)P^*_{n'k'}(132)}
{
	(\omega_{ki}-\omega)(\omega_{k'n'}-\omega)
}
+
\frac{P_{nk}(132)P^*_{n'k'}(132)}
{
	(\omega_{kn}-\omega)(\omega_{k'n'}-\omega)
}
\right\rbrace
.
\end{eqnarray}
\end{widetext}

Summation over projections can be performed separately for each term. Using Eq. (\ref{cyclic}) and the Eckart-Wigner theorem for the first summand in curly brackets in Eq. (\ref{s2}) we obtain
\begin{align}
\label{sb3}
&
P_{nk}(123)P^*_{n'k'}(123)=
\sum\limits_{\substack{q_1q_2q_3\\q_1'q_2'q_3'}}
(-1)^{\phi+F_{i}+F_{f}+F_{n}+F_{n'}}
\\\nonumber
\times
&
(-1)^{F_{k}+F_{k'}-M_{F_i}-M_{F_f}-M_{F_{n}}-M_{F_{n'}}-M_{F_{k}}-M_{F_{k'}}}
\\\nonumber
\times
&
\begin{pmatrix}
   F_f & 1 & F_n \\
   -M_{F_f} & -q_1 & M_{F_n}
\end{pmatrix}
\begin{pmatrix}
   F_n & 1 & F_{k} \\
   -M_{F_n} & -q_2 & M_{F_{k}}
\end{pmatrix}
\\\nonumber
\times
&
\begin{pmatrix}
   F_{k} & 1 & F_i \\
   -M_{F_{k}} & -q_3 & M_{F_i}
\end{pmatrix}
\begin{pmatrix}
   F_i & 1 & F_{k'} \\
   -M_{F_i} & -q_3' & M_{F_{k'}}
\end{pmatrix}
\\\nonumber
\times
&
\begin{pmatrix}
   F_{k'} & 1 & F_{n'} \\
   -M_{F_{k'}} & -q_2' & M_{F_{n'}}
\end{pmatrix}
\begin{pmatrix}
   F_{n'} & 1 & F_f \\
   -M_{F_{n'}} & -q_1' & M_{F_f}
\end{pmatrix}
\\\nonumber
\times
&
(e_1^*)_{q_1}(e_2)_{q_2}(e_3)_{q_3}
%\\\nonumber
%\times
%&
\left[(e_3^*)_{q'_3}(e_2^*)_{q'_2}(e_1)_{q'_1}\right]
P_{\mathrm{red}}
&
,
\end{align}
where 
%\begin{eqnarray}
%\nonumber
$\phi=q_1+q_2+q_3+q_1'+q_2'+q_3'$
%\end{eqnarray}
and $ P_{\mathrm{red}} $ denotes the product of reduced matrix elements 
\begin{eqnarray}
\label{pred}
\left.
\begin{aligned}
P_{\mathrm{red}}=
&
\langle n_{f}l_{f}j_{f}F_{f}|| r || nl_{n}j_{n}F_{n}\rangle
\langle nl_{n}j_{n}F_{n}|| r ||kl_{k}j_{k}F_{k}\rangle
\\\nonumber
\times
&
\langle kl_{k}j_{k}F_{k}|| r || n_{i}l_{i}j_{i}F_{i}\rangle
\langle n_{i}l_{i}j_{i}F_{i}||r ||  kl_{k'}j_{k'}F_{k'} \rangle
\\\nonumber
\times
&
\langle k'l_{k'}j_{k'}F_{k'}|| r ||n'l_{n'}j_{n'}F_{n'}\rangle
\\\nonumber
\times
&
\langle n'l_{n'}j_{n'}F_{n'}|| r ||  n_{f}l_{f}j_{f}F_{f}\rangle
\end{aligned}
\right.
.
\end{eqnarray} 
Then summation over projections corresponding to the inital ($ M_{F_i} $), final ($ M_{F_f} $) and intermdiate ($ M_{F_n} $, $ M_{F_{n'}} $, $ M_{F_k} $, $ M_{F_{k'}} $) states in Eq. (\ref{s2}) can be done using equality (see Eq. (29) page 392 in \cite{VMK}):
\begin{align}
\label{s3}
&
\sum\limits_{\substack{\mathrm{all}\\\mathrm{projections} }}
P_{nk}(123)P^*_{n'k'}(123)=
\sum\limits_{\substack{q_1q_2q_3\\q_1'q_2'q_3'}}
\sum_{\substack{xyz \\ \xi\eta\zeta}}\Pi^2_{xyz}
\\\nonumber
\times
&
(-1)^{\phi}
\begin{pmatrix}
   1   & 1   & x \\
   q_1 & q_2 & \xi
\end{pmatrix}
\begin{pmatrix}
   1   & 1    & y \\
   q_3 & q'_3 & \eta
\end{pmatrix}
\begin{pmatrix}
   1 & 1 & z \\
   q'_2 & q'_1 & \zeta
\end{pmatrix}
\\\nonumber
\times
&
\begin{pmatrix}
   x & y & z \\
   -\xi & -\eta & -\zeta
\end{pmatrix}
\begin{Bmatrix}
   1 & x & 1 \\
   F_k & F_n & F_f
\end{Bmatrix}
\begin{Bmatrix}
   1 & y & 1 \\
   F_{k'} & F_i & F_k
\end{Bmatrix}
\\\nonumber
\times
&
\begin{Bmatrix}
   1 & z & 1 \\
   F_f & F_{n'} & F_{k'}
\end{Bmatrix}
\begin{Bmatrix}
   x & y & z \\
   F_{k'} & F_f & F_k
\end{Bmatrix}
\\\nonumber
\times
&
(e_1^*)_{q_1}(e_2)_{q_2}(e_3)_{q_3}
\left[(e_3^*)_{q'_3}(e_2^*)_{q'_2}(e_1)_{q'_1}\right]
P_{\mathrm{red}}.
&
\end{align}

Remaining summation over $ q_i $ and $ q_j' $ ($ i,\,j=1,\,2,\,3 $) in Eq. (\ref{s3}) with the use of Eqs. (\ref{r2}) and (\ref{r3}) yields 
\begin{align}
\label{s4}
&
\sum\limits_{\substack{\mathrm{all}\\\mathrm{projections} }}
P_{nk}(123)P^*_{n'k'}(123)=
\sum_{\substack{xyz \\ \xi\eta\zeta}}(-1)^{y-\eta}
\\\nonumber
\times
&
(-1)^{x+z+\eta}
\Pi_{xyz} 
\lbrace e_{1}^*\otimes e_{2} \rbrace_{x\xi}
\lbrace e_{2}^*\otimes e_{1}\rbrace_{z\zeta}
\lbrace e_{3}\otimes e_{3}^*\rbrace_{y-\eta}
P_{\mathrm{red}}
\\\nonumber
\times
&
\begin{pmatrix}
   x & z & y \\
   \xi & \zeta & -\eta
\end{pmatrix}
\begin{Bmatrix}
   1 & x & 1 \\
   F_k & F_n & F_f
\end{Bmatrix}
\begin{Bmatrix}
   1 & y & 1 \\
   F_{k'} & F_i & F_k
\end{Bmatrix}
\\\nonumber
\times
&
\begin{Bmatrix}
   1 & z & 1 \\
   F_f & F_{n'} & F_{k'}
\end{Bmatrix}
\begin{Bmatrix}
   x & y & z \\
   F_{k'} & F_f & F_k
\end{Bmatrix}.
\end{align}
Then, by the definition of the irreducible tensor product, the expression (\ref{s4}) recasts to
%\newpage
\begin{eqnarray}
\label{s5}
\sum\limits_{\substack{\mathrm{all}\\\mathrm{projections} }}
P_{nk}(123)P^*_{n'k'}(123)=
\sum_{\substack{xyz \\ \eta}}(-1)^{y-\eta}
\\\nonumber
\times
\Pi_{xz}
\begin{Bmatrix}
   1 & x & 1 \\
   F_k & F_n & F_f
\end{Bmatrix}
\begin{Bmatrix}
   1 & y & 1 \\
   F_{k'} & F_i & F_k
\end{Bmatrix}
\\\nonumber
\times
\begin{Bmatrix}
   1 & z & 1 \\
   F_f & F_{n'} & F_{k'}
\end{Bmatrix}
\begin{Bmatrix}
   x & y & z \\
   F_{k'} & F_f & F_k
\end{Bmatrix}
\\\nonumber
\times
\lbrace
\lbrace e_{1}^*\otimes e_{2} \rbrace_{x}
\otimes
\lbrace e_{2}^*\otimes e_{1}\rbrace_{z}
\rbrace
_{y\eta}
%\\\nonumber
%\times
\lbrace e_{3}\otimes e_{3}^*\rbrace_{y-\eta}
P_{\mathrm{red}}
.
\end{eqnarray}

According to Eq. (\ref{r4}) the sum over $ \eta $ in Eq. (\ref{s5}) could be rewritten as the scalar product of two tensors of rank $ y $
\begin{eqnarray}
\label{s6}
\sum\limits_{\substack{\mathrm{all}\\\mathrm{projections} }}
P_{nk}(123)P^*_{n'k'}(123)=
\sum_{xyz}\Pi_{xz}
\\\nonumber
\times
\begin{Bmatrix}
   1 & x & 1 \\
   F_k & F_n & F_f
\end{Bmatrix}
\begin{Bmatrix}
   1 & y & 1 \\
   F_{k'} & F_i & F_k
\end{Bmatrix}
\\\nonumber
\times
\begin{Bmatrix}
   1 & z & 1 \\
   F_f & F_{n'} & F_{k'}
\end{Bmatrix}
\begin{Bmatrix}
   x & y & z \\
   F_{k'} & F_f & F_k
\end{Bmatrix}
\\\nonumber
\times
\lbrace
\lbrace e_{1}^*\otimes e_{2} \rbrace_{x}
\otimes
\lbrace e_{2}^*\otimes e_{1}\rbrace_{z}
\rbrace
_{y}
\\\nonumber
\cdot
\lbrace e_{3}\otimes e_{3}^*\rbrace_{y}
P_{\mathrm{red}}
,
\end{eqnarray}
or, in equivalent form,
\begin{eqnarray}
\label{s6a}
\sum\limits_{\substack{\mathrm{all}\\\mathrm{projections} }}
P_{nk}(123)P^*_{n'k'}(123)=
\sum_{xyz}\Pi_{xyz}
\\\nonumber
\times
\begin{Bmatrix}
   1 & x & 1 \\
   F_k & F_n & F_f
\end{Bmatrix}
\begin{Bmatrix}
   1 & y & 1 \\
   F_{k'} & F_i & F_k
\end{Bmatrix}
\\\nonumber
\times
\begin{Bmatrix}
   1 & z & 1 \\
   F_f & F_{n'} & F_{k'}
\end{Bmatrix}
\begin{Bmatrix}
   x & y & z \\
   F_{k'} & F_f & F_k
\end{Bmatrix}
\\\nonumber
\times
\lbrace
\lbrace
\lbrace e_{1}^*\otimes e_{2} \rbrace_{x}
\otimes
\lbrace e_{2}^*\otimes e_{1}\rbrace_{z}
\rbrace
_{y}
%\right.
\\\nonumber
%\left.
\otimes
\lbrace e_{3}\otimes e_{3}^*\rbrace_{y}
\rbrace_{00}
P_{\mathrm{red}}
.
\end{eqnarray}
Conversing the coupling scheme in the tensor product in Eq. (\ref{s6a}) with the use of Eq. (\ref{r5a}), we get
\begin{eqnarray}
\label{o1}
\sum\limits_{\substack{\mathrm{all}\\\mathrm{projections} }}
P_{nk}(123)P^*_{n'k'}(123)
=
\sum_{xyzg}(-1)^{\psi}
\\\nonumber
\times
\Pi_{xz}
\Pi_{y}^2
\begin{Bmatrix}
x & z & y \\
1 & 1 & g
\end{Bmatrix}
\begin{Bmatrix}
   1 & x & 1 \\
   F_k & F_n & F_f
\end{Bmatrix}
\\\nonumber
\times
\begin{Bmatrix}
   1 & y & 1 \\
   F_{k'} & F_i & F_k
\end{Bmatrix}
\begin{Bmatrix}
   1 & z & 1 \\
   F_f & F_{n'} & F_{k'}
\end{Bmatrix}
\\\nonumber
\times
\begin{Bmatrix}
   x & y & z \\
   F_{k'} & F_f & F_k
\end{Bmatrix}
%\\\nonumber
\times
U_{xzg}
P_{\mathrm{red}}\equiv O^{123}_{123}
,
\end{eqnarray}
where $ \psi = y+g+1 $, and the tensor product $U_{xzg}$ is
%\begin{widetext}
\begin{eqnarray}
\label{Uxzg}
U_{xzg}\equiv
\lbrace
\lbrace e_{1}^*\otimes e_{2} \rbrace_{x}
\otimes
e_{3}
\rbrace
_{g}
%\\\nonumber
\cdot
\lbrace
\lbrace e_{1}\otimes e_{2}^* \rbrace_{z}
\otimes e_{3}^*
\rbrace_{g}.\qquad
\end{eqnarray}
%\end{widetext}
All the necessary information about the angular correlations is contained in the expression above. 

An explicit dependence on the angle between polarizations arises from the relations:
\begin{gather}
\label{udefmain}
U_{001}=\frac{\cos^2{\theta_{12}}}{3},\\
\nonumber
U_{110}=-\frac{1}{6}\sin^2{\theta_{12}}\cos^2{\theta_{123}},\\
\nonumber
U_{111}=\frac{1}{4}(\cos^2{\theta_{13}} - 2\cos{\theta_{12}}\cos{\theta_{13}}\cos{\theta_{23}}\\
\nonumber
 + \cos^2{\theta_{23}}),\\
\nonumber
U_{221}=\frac{1}{60}(4\cos^2{\theta_{12}} + 9\cos^2{\theta_{13}} + 9\cos^2{\theta_{23}}
\\\nonumber
 - 6\cos{\theta_{12}}\cos{\theta_{13}}\cos{\theta_{23}}),\\
 \nonumber
U_{021}=U_{201}=\frac{1}{6\sqrt{5}}(6\cos{\theta_{12}}\cos{\theta_{13}}\cos{\theta_{23}} \\
\nonumber
- 2\cos^2{\theta_{12}}),\\
\nonumber
U_{121}=U_{211}=-\frac{1}{4\sqrt{15}}(3\cos^2{\theta_{23}} - 3\cos^2{\theta_{13}})
,
\end{gather}
where $ \theta_{12} $ is the angle between vectors $ \textbf{e}_{1}$ and $ \textbf{e}_{2}$,  $ \theta_{13} $ is the angle between vectors $ \textbf{e}_{1}^*$ and $ \textbf{e}_{3}$, $ \theta_{23} $ is the angle between vectors $ \textbf{e}_{2}$ and $ \textbf{e}_{3}$ and $\theta_{123}$ is the angle between vector product $[\textbf{e}_1^*\times \textbf{e}_2]$ and $\textbf{e}_3$.

In the case of two parallel polarizations of incident laser photons, when the vector $ \textbf{e}_{2}$ is parallel to the vector $ \textbf{e}_3 $ and, therefore, $ \theta_{23}=0 $ and $ \theta_{12}=\theta_{13}\equiv \theta $, the following nonzero contributions of $ U_{xzg} $ are
\begin{eqnarray}
\label{udef}
\begin{aligned}
&
U_{001}=\frac{\cos ^2(\theta )}{3},\\
&
U_{111}=\frac{\sin ^2(\theta )}{4},\\
&
U_{221}=\frac{7 \cos ^2(\theta )}{60}+\frac{3}{20},\\
&
U_{021}=U_{201}=\frac{2 \cos ^2(\theta )}{3 \sqrt{5}},\\
&
U_{121}=U_{211}=-\frac{1}{4} \sqrt{\frac{3}{5}} \sin ^2(\theta )
.
&
\end{aligned}
\end{eqnarray}
%Here $ \theta\equiv\theta_{12} $ is the angle between vectors $ \textbf{e}_{1}^* $ and $ \textbf{e}_{2} $ (or $ \textbf{e}_3 $). 

Summation over projections in the remaining three terms in Eq. (\ref{s2}) can be performed in the same way. This yields
\begin{eqnarray}
\label{o2}
\sum\limits_{\substack{\mathrm{all}\\\mathrm{projections} }}
P_{nk}(123)P^*_{n'k'}(132)
=
\sum_{xyzgh'}(-1)^{\psi}
\\\nonumber
\times
\Pi_{xz}
\Pi_{y}^2
(-1)^{z+h'}\Pi_{zh'}
\begin{Bmatrix}
x & z & y \\
1 & 1 & g
\end{Bmatrix}
\begin{Bmatrix}
1 & 1 & z \\
g & 1 & h'
\end{Bmatrix}
\\\nonumber
\times
\begin{Bmatrix}
   1 & x & 1 \\
   F_k & F_n & F_f
\end{Bmatrix}
\begin{Bmatrix}
   1 & y & 1 \\
   F_{k'} & F_i & F_k
\end{Bmatrix}
\\\nonumber
\times
\begin{Bmatrix}
   1 & z & 1 \\
   F_f & F_{n'} & F_{k'}
\end{Bmatrix}
\begin{Bmatrix}
   x & y & z \\
   F_{k'} & F_f & F_k
\end{Bmatrix}
\\\nonumber
\times
U_{xh'g}
P_{\mathrm{red}}\equiv O^{123}_{132}
,
\end{eqnarray}
\begin{eqnarray}
\label{o3}
\sum\limits_{\substack{\mathrm{all}\\\mathrm{projections} }}
P_{nk}(132)P^*_{n'k'}(123)
=
\sum_{xyzgh}(-1)^{\psi}
\\\nonumber
\times
\Pi_{xz}
\Pi_{y}^2
(-1)^{x+h}\Pi_{xh}
\begin{Bmatrix}
x & z & y \\
1 & 1 & g
\end{Bmatrix}
\begin{Bmatrix}
1 & 1 & x \\
g & 1 & h
\end{Bmatrix}
\\\nonumber
\times
\begin{Bmatrix}
   1 & x & 1 \\
   F_k & F_n & F_f
\end{Bmatrix}
\begin{Bmatrix}
   1 & y & 1 \\
   F_{k'} & F_i & F_k
\end{Bmatrix}
\\\nonumber
\times
\begin{Bmatrix}
   1 & z & 1 \\
   F_f & F_{n'} & F_{k'}
\end{Bmatrix}
\begin{Bmatrix}
   x & y & z \\
   F_{k'} & F_f & F_k
\end{Bmatrix}
\\\nonumber
\times
U_{hzg}
P_{\mathrm{red}}\equiv O^{132}_{123}
,
\end{eqnarray}
\begin{eqnarray}
\label{o4}
\sum\limits_{\substack{\mathrm{all}\\\mathrm{projections} }}
P_{nk}(132)P^*_{n'k'}(132)
=
\sum_{xyzghh'}(-1)^{\psi}
\\\nonumber
\times
\Pi_{xz}
\Pi_{y}^2
(-1)^{z+x+h+h'}\Pi_{xzhh'}
\begin{Bmatrix}
x & z & y \\
1 & 1 & g
\end{Bmatrix}
\\\nonumber
\times
\begin{Bmatrix}
1 & 1 & z \\
g & 1 & h'
\end{Bmatrix}
\begin{Bmatrix}
1 & 1 & x \\
g & 1 & h
\end{Bmatrix}
\\\nonumber
\times
\begin{Bmatrix}
   1 & x & 1 \\
   F_k & F_n & F_f
\end{Bmatrix}
\begin{Bmatrix}
   1 & y & 1 \\
   F_{k'} & F_i & F_k
\end{Bmatrix}
\\\nonumber
\times
\begin{Bmatrix}
   1 & z & 1 \\
   F_f & F_{n'} & F_{k'}
\end{Bmatrix}
\begin{Bmatrix}
   x & y & z \\
   F_{k'} & F_f & F_k
\end{Bmatrix}
\\\nonumber
\times
U_{hh'g}
P_{\mathrm{red}}\equiv O^{132}_{132}
,
\end{eqnarray}
where we used Eq. (\ref{r5}). Then substitution of Eqs. (\ref{o1}), (\ref{o2}), (\ref{o3}) and (\ref{o4}) into Eq. (\ref{s2}) leads to
\begin{gather}
\label{theend}
\sum\limits_{M_{F_i}M_{F_f}}T_{n}T_{n'}^*=
\omega^3\omega_{nf}^{3/2}\omega_{n'f}^{3/2}
\\\nonumber
\times
\left(\omega_{ni}-\omega \right)^{3/2}
\left(\omega_{n'i}-\omega \right)^{3/2}
\sum\limits_{\substack{kl_kj_kF_k\\k'l_{k'}j_{k'}F_{k'}}}
\\\nonumber
\times
\left\lbrace
\frac{O^{123}_{123}}
{
	(\omega_{ki}-\omega)(\omega_{k'i}-\omega)
}
+
\frac{O^{132}_{123}}
{
	(\omega_{kn}+\omega)(\omega_{k'i}-\omega)
}
\right.
\\\nonumber
+
\left.
\frac{O^{123}_{132}}
{
	(\omega_{ki}-\omega)(\omega_{k'n'}-\omega)
}
+
\frac{O^{132}_{132}}
{
	(\omega_{kn}-\omega)(\omega_{k'n'}-\omega)
}
\right\rbrace
.
\end{gather}

The expression (\ref{theend}) contains all necessary angular correlations for the three-photon scattering cross section given by Eqs. (\ref{sigmamaintext})-(\ref{fnr}) in the main text.Summing over quantum numbers $ l_kj_kF_k$ and $l_{k'}j_{k'}F_{k'} $ in (\ref{theend}) numerically, one can obtain the common angular depended factor for the interfering contributions, Eq. (\ref{theend}). This factor arises finally in the expressions for the NR corrections, see the main text.

\appendix
\renewcommand{\theequation}{B\arabic{equation}}
\setcounter{equation}{0}
\section*{Appendix B: Derivation of NR corrections}

%This Appendix provides general expressions for the NR corrections discussed in sections \ref{section2} and \ref{section3}. 
The NR correction to $ 2s_{1/2}^{F=1}\rightarrow ns_{1/2}^{F=1} $ (with $ n=4,\,6,\,8,\,12 $) transition frequencies Eq. (\ref{nrsum2}) occurs using Eqs. (\ref{theend}), (\ref{fres}) and (\ref{fnr}). Then summing over quantum numbers $ j_{f}F_{f} $ and $ l_{k}j_{j}F_{k} $ in Eq. (\ref{nrsum2}), we arrive at
\begin{widetext}
\begin{eqnarray}
\label{nrsum3}
\delta_{\mathrm{NR}}(2s_{1/2}^{F=1}-ns_{1/2}^{F=1} )=
\frac{\Gamma_{ns_{1/2}}^2}{4}
\left(
\frac{1}{25\Delta_{1}}
%\right.
%\\\nonumber
%\left.
+
\frac{1}{25\Delta_{2}}
+
\frac{2}{75\Delta_{3}}
+
\frac{7}{75\Delta_{4}}
\right)
\frac{\beta_{2pnd2s}(\omega_{ns2s}/2)}{\beta_{2pns2s}(\omega_{ns2s}/2)}
\\
\nonumber
\times
\frac{10 U_{001}+\sqrt{5} U_{021}-15 U_{111}-\sqrt{15} U_{121}+10 \sqrt{5} U_{201}+5 \sqrt{15} U_{211}+5 U_{221}}{ U_{001}+\sqrt{5} U_{021}+3 U_{111}-\sqrt{15}
U_{121}+\sqrt{5} U_{201}-\sqrt{15} U_{211}+5 U_{221}}
.
\end{eqnarray} 
\end{widetext}
Here the coefficient $ \beta $ in Eq. (\ref{nrsum3}) is defined by
\begin{eqnarray}
\label{beta}
\beta_{2pns(nd)2s}(\omega)=
I_{2pns}
\sum\limits_{k}
\left\lbrace
\frac{
	I_{ns(nd)kp}
	I_{kp2s}
}
{
	E_{kp}-E_{2s}-\omega
}
\right.
\\\nonumber
+
\left.
\frac{
	I_{ns(nd)kp}
	I_{kp2s}
}
{
	E_{kp}-E_{ns(nd)}+\omega
}
\right\rbrace
,
%\\
\end{eqnarray}
%where 
\begin{eqnarray}
I_{n'l'nl}=\int\limits_{0}^{\infty}r^3R_{n'l'}R_{nl}dr
\end{eqnarray}
and $ R_{nl} $ represents the corresponding radial part of the Schr\"{o}dinger wave function, Summation over $k$ runs all the entire spectrum including the continuum. 
The numerical values of Eq. (\ref{beta}) calculated by the Green's function method are listed in Table~\ref{tab1}. For the particular case of parallel (anti-parallel) polarizations of incident photons, the relations Eqs. (\ref{udef}) should be used. Then, substituting Eqs. (\ref{udef}) into Eq. (\ref{nrsum3}), we obtain Eqs. (\ref{f4})-(\ref{f12}).

After performing a similar calculation, the NR correction to $ 2s_{1/2}^{F=0}-ns_{1/2}^{F=0} $  ($ n=4,\,6,\,8,\,12 $) transition frequencies Eq. (\ref{nrsum4}) can be found as
\begin{widetext}
\begin{eqnarray}
\label{nrsum5}
\delta_{\mathrm{NR}}(2s_{1/2}^{F=0}-ns_{1/2}^{F=0} )=
\frac{\Gamma_{ns_{1/2}}^2}{4}
\left(
\frac{2}{25\Delta_{1}'}
%\right.
%\\\nonumber
%\left.
+
\frac{3}{25\Delta_{2}'}
\right)
\frac{\beta_{2snd2p}(\omega_{ns2s}/2)}{\beta_{2sns2p}(\omega_{ns2s}/2)}
\\
\nonumber
\times
\frac{10 U_{001}+\sqrt{5} U_{021}-15 U_{111}-\sqrt{15} U_{121}+10 \sqrt{5} U_{201}+5 \sqrt{15} U_{211}+5 U_{221}}{U_{001}+\sqrt{5} U_{021}+3 U_{111}-\sqrt{15} U_{121}+\sqrt{5} U_{201}-\sqrt{15} U_{211}+5 U_{221}}
.
\end{eqnarray} 
\end{widetext}
Note that the correlation factor here coincides with Eq. (\ref{nrsum3}).

The situation is different for the NR correction to the $ 2s_{1/2}^{F=1}-nd_{3/2}^{F=2} $ ($ n=4,\,6,\,8,\,12 $) transition frequencies, see Eq. (\ref{nrsumnd1}):
\begin{eqnarray}
\label{nrsumnd2}
\delta_{\mathrm{NR}}(2s_{1/2}^{F=1}-nd_{3/2}^{F=2})=-\frac{\Gamma_{nd_{3/2}}^2}{4\Delta''}
\frac{1}{11}
.
\end{eqnarray} 
This correction is independent of angles, and, therefore, can not be eliminated by choosing the geometry of the experiment.

The remaining NR correction to $ 2s_{1/2}^{F=1}-nd_{3/2}^{F=1} $ ($ n=4,\,6,\,8,\,12 $) transition frequencies due to the neighbouring $ ns_{1/2}^{F=1} $ state (see Eq. (\ref{nrsumnd22})) is
\begin{widetext}
\begin{eqnarray}
\label{nrsumnd3}
\delta_{\mathrm{NR}}(2s_{1/2}^{F=1}-nd_{3/2}^{F=1} )=
\frac{\Gamma_{nd_{3/2}}^2}{4}
\left(\frac{50}{33\Delta'''}
\right)
\frac{\beta_{2sns2p}(\omega_{2snd}/2)}{\beta_{2snd2p}(\omega_{2snd}/2)}
\\\nonumber
\times
\frac{10 U_{001}+10 \sqrt{5} U_{021}-15 U_{111}+5 \sqrt{15} U_{121}+\sqrt{5} U_{201}-\sqrt{15} U_{211}+5 U_{221}}{20 U_{001}+2 \sqrt{5} U_{021}+15 U_{111}+\sqrt{15} U_{121}+2 \sqrt{5}
   U_{201}+\sqrt{15} U_{211}+U_{221}}
.
\end{eqnarray} 
\end{widetext}

%Derived above Eqs. (\ref{nrsum3}), (\ref{nrsum5}), (\ref{nrsumnd2}) and (\ref{nrsumnd3}) fully describe spatial behaviour of leading order NR correction to transition frequencies considered in the present paper. 
\begin{table}
\caption{Coefficients $ \beta $ in a.u. and level widths $ \Gamma_{nlj} $ in MHz.}
\begin{tabular}{c c c c c}
\hline
\hline
$n$ & $ \beta_{2sns2p} $ & $ \beta_{2snd2p} $ & $ \Gamma_{ns_{1/2}} $ & $ \Gamma_{nd_{3/2}} $  \\
\hline
4  & -38.1593 & 2449.09 & 0.70 & 4.41\\
6  &  13.949  & 591.154 & 0.29 & 1.33\\
8  &  8.41272 & 240.557 & 0.14 & 0.56\\
12 &  3.10962 & 71.980 & 0.05 & 0.17\\
\hline
$n$ & $ \beta_{1sns2p} $ & $ \beta_{1snd2p} $ & $ \Gamma_{ns_{1/2}} $  & $ \Gamma_{nd_{3/2}} $  \\
\hline
3  & 187.375 & 1.005 & 1.01 &  10.30\\
\hline
\end{tabular} 
\label{tab1}
\end{table}

\bibliography{mybibfile} 

\end{document}